\documentstyle[12pt,cite,epsfig,psfrag]{article}

\def\be{\begin{equation}}
\def\ee{\end{equation}}
\def\bea{\begin{eqnarray}}
\def\eea{\end{eqnarray}}

\newcommand{\beqal}{\begin{eqnarray}\label}
\newcommand{\beqa}{\begin{eqnarray}}
\newcommand{\eeqa}{\end{eqnarray}}

\begin{document}

\begin{titlepage}
\begin{center}
%\hfill hep-th/yymmnnn\\
%\hfill IP/BBSR/2008-00\\

\vskip .2in

{\Large \bf Black hole phase transitions via Bragg-Williams}
\vskip .5in

{\bf Souvik Banerjee\footnote{e-mail: souvik@iopb.res.in}
\bf Sayan K. Chakrabarti\footnote{e-mail: sayan@iopb.res.in}, 
\bf Sudipta Mukherji\footnote{e-mail: mukherji@iopb.res.in},\\
\bf Binata Panda\footnote{e-mail: binata@iopb.res.in}\\
\vskip .1in
{\em Institute of Physics,\\
Bhubaneswar 751~005, India.}}
\end{center}

\begin{center} {\bf ABSTRACT}

\end{center}
\begin{quotation}\noindent
\baselineskip 15pt

We argue that a convenient way to analyze instabilities of black holes in AdS 
space is via Bragg-Williams construction of a free energy function. Starting with a 
pedagogical review of this construction in condensed matter systems and also its implementation to 
Hawking-Page transition, we study instabilities associated with hairy black holes and also 
with the $R$-charged black holes. For the hairy black holes, an analysis of thermal quench
is presented.

\end{quotation}
\vskip 2in
December 2010\\
\end{titlepage}
\vfill
\eject

%\begin{quote}
%\noindent
%\end{quote}
%\setcounter{footnote}{0}
\section{Introduction}

Within the mean field approximation, phase transition is primarily described 
via Landau theory. Under the assumptions that the order parameter is
small and uniform near the transition, this theory provides us with a wealth of
information about the nature of the phase transition. It is 
based upon a power series expansion of free energy in terms of the order
parameter. The terms in this expansion are normally determined by symmetry 
considerations of the phases. Furthermore, owing to the smallness of
the order parameter, only a few leading terms are kept.  The usefulness of 
the Landau theory lies in its simplicity as most of its predictions 
can be achieved by solving simple algebraic equations \cite{Chaikin}.
While this theory is most suitable in describing a second order phase transition,
one needs to be somewhat careful to treat first order phase transition within this 
framework. This is because, in a first order transition, order parameter suffers
a discontinuous jump across the critical temperature. If this change is large,
a power series expansion of free energy may acquire ambiguities. One then
requires a more complete mean field theory. An example of this kind is the
Bragg-Williams (BW) theory \cite{bw1, bw2}. Originally used to describe order - disorder 
transition of alloys, it has a wide range of applications \cite{Chaikin, kubo}.
In this approach, one constructs an approximate expression for the free energy in 
terms of the order parameter and uses the condition that its equilibrium value minimizes the 
free energy. Our aim in this paper is to use this approach to study 
phase transition involving black holes in the presence of a negative 
cosmological constant.

\begin{figure}[t]
\begin{center}
\begin{psfrags}
\psfrag{b}[][]{$\bar {T}$}
\psfrag{a}[][]{$\bar \mu$}
\psfrag{c}[][]{$\bar T = \frac{1}{\pi}$}
%\psfrag{d}[][]{IV}
%\psfrag{3/4}[][]{$3/4$}
%\psfrag{0.03}[][]{$0.03$}
\epsfig{file=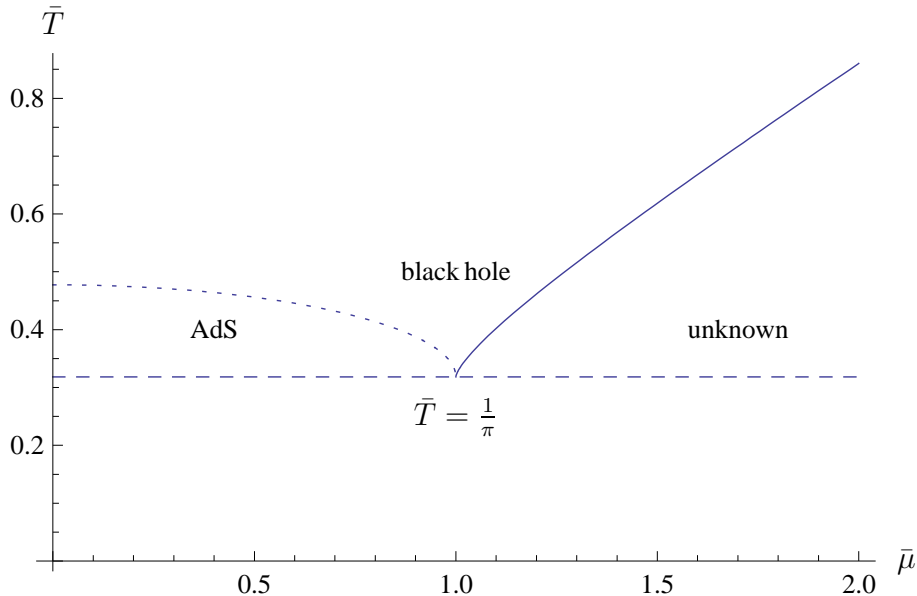, width= 12cm,angle=0}
\end{psfrags}
\vspace{ .1 in }
\caption{Phase diagram for the $R$-charged black hole with single charge shown in temperature, chemical potential plane.
Line separating thermal AdS and black hole represents the first order
phase
transition line given by equation (\ref{forder}). On the other hand, the
line between black hole and the unknown phase is a second order line - the
equation of which is given in (\ref{sorder}). The dashed line is for $\bar
T
= 1/\pi$ below which we can not extend various phases.}
\label{phased}
\end{center}
\end{figure}

In 1970, following the work of Bekenstein and others, Hawking showed that stationary
black holes have temperature $T$ related to their  surface gravity, and, they
behave like thermodynamic objects. In the presence of a negative 
cosmological constant, these black holes asymptote to the anti-de Sitter (AdS)
space and, provided that their size is {\it sufficiently large}, the 
temperature typically increases with their internal energy. Therefore, unlike black holes
in Minkowski space, these have positive specific heat and hence they are
thermodynamically stable. However, as we reduce the temperature, at some
stage, various instabilities creep in. A prime example of
this kind is the well known Hawking-Page (HP) instability \cite{Hawking:1982dh}, where below a critical
temperature, a AdS-Schwarzschild black hole becomes unstable and crosses over
to the thermal AdS space via a first order phase transition. A similar behaviour arises
for the Reissner-Nodstr\"om black holes in AdS space in the grand canonical 
ensemble \cite{Peca:1998cs, Chamblin:1999tk}. Subsequently, it was found 
that the charged black holes in five 
dimensional 
${\cal{N}} = 2$ gauged  supergravity theory also exhibit rich phase structures. Black holes in this 
theory, known as $R$-charged black holes,  can carry  three independent gauge charges and the stability of 
these black holes were studied, for example,
in \cite{Gubser:1998jb, Cvetic:1999ne, Yamada:2007gb}. For single $R$-charged holes, the phase
structure is shown in figure \ref{phased}. It is plotted in the $T - \mu$ plane where $\mu$ is 
the chemical 
potential conjugate to the charge.
There are three distinct phases, namely, the thermal AdS, black hole and a yet unknown phase. At 
a low 
temperature and small chemical potential, the system
is always in thermal AdS phase. The cross-over from AdS to the black hole phase is shown by the
dotted line in the plot. This is the usual first order HP transition.
The black hole phase at fixed temperature also becomes unstable once the
chemical potential is increased beyond a critical value. The corresponding stable
phase is unknown as yet\footnote{It may also be possible that there is no stable phase
at all.}. However, if a stable phase exists, this transition would be a continuous phase
transition marked by divergences of specific heat and susceptibility. The solid line
in figure \ref{phased} represents this critical line. 
Another interesting example of phase transition in black hole physics includes
holes with scalar hair. In AdS space, it is possible to construct black holes 
with hair. The one which we consider here was found in \cite{Martinez:2005di}. 
These are electrically charged black hole solutions in four dimensional AdS space
with a conformally coupled scalar. Unlike previous examples, here, the horizon
is a negatively curved two dimensional constant curvature manifold. We will call
these holes as hairy black holes in this paper. 
A recent analysis on the stability of these black holes were carried 
out in \cite{Martinez:2010ti}. It was concluded that
at low temperature, the hairy black hole
is stable. At high temperature, by losing ``hair", it becomes unstable and crosses over to a 
Reissner-Nodstr\"om phase with same horizon topology. 
%\begin{figure}[h]
%\begin{center}
%\begin{psfrags}
%\psfrag{b}[][]{$\bar {T}$}
%\psfrag{a1}[][]{$a$}
%\psfrag{c}[][]{$\bar T = \frac{1}{\pi}$}
%\psfrag{d}[][]{IV}
%\psfrag{3/4}[][]{$3/4$}
%\psfrag{0.03}[][]{$0.03$}
%\epsfig{file=tvsa1.eps, width= 12cm,angle=0}
%\end{psfrags}
%\vspace{ .1 in }
%\caption{Phase diagram for the hairy black hole in $\bar T$ - $a$ plane. The upper and lower 
%dashed lines represent $\bar T_{\rm{max}}$ and $\bar T_{\rm{min}}$ respectively. The dotted line 
%is for $\bar T=\frac{1}{2\pi}$. The dash-dotted line shows the first order transition and the 
%thick dotted line at the top denotes metastability. The solid vertical line is like a cut-off line 
%below which one can not predict about the phase structure. } 
%\label{phase_hairy}
%\end{center}
%\end{figure}
Our aim in this paper is to study various instabilities mentioned  
in the previous paragraph within the framework of BW theory. As we will see, 
all these transitions are very elegantly captured within this scheme. Its implementation
is simple, it makes definitive statements about the nature of the phase transition and
it gives us the mean field critical exponents where the black hole undergoes a continuous
transition. 
In particular, for hairy to Reissner-Nordstr\"om transition, we construct the
off-shell free energy function in terms of a suitably chosen order parameter
and study the behaviour  around its saddle points. 
We further use the BW potential to analyze the system under temperature quench. It turns out that 
the time variation of the order parameter after quench from unstable to its stable minimum can
be semi-analytically constructed.
As for the
$R$-charged black holes, we reproduce the complete phase diagram in figure \ref{phased} 
and also compute
the classical critical exponents near the second order instability.

Our motivation
to analyse phases via BW theory grew out of gauge/gravity correspondence.  
Within this correspondence, gravity theory in AdS space is expected to have
a gauge theory dual on the boundary. Details of these
gauge theories depend on how one embeds AdS in ten or eleven dimensions, nature
of the compactifications etcetera, and, in many cases, are not explicitly known. 
However, the dual nature of the correspondence suggests that since the 
gravity is weakly coupled, the gauge theory, if exists,  has to have a strong coupling. 
Due to the lack of a systematic approach to handle strongly coupled theories, 
direct gauge theoretic computations become difficult. However, the behaviour of the weakly coupled
gravity along with gauge/gravity duality often helps us exploring strongly coupled
gauge theories. Association of deconfining transition of large $N$, ${\cal N} =4$
Yang-Mills with the HP transition in five dimensional AdS space is a classic example 
in this regard \cite{Witten:1998zw}. Our hope is to construct candidate effective 
potentials for gauge theories describing various phases at non-zero temperature 
via computations of BW potentials from their gravity duals. For a partial success in
this direction for $R$-charged black hole with flat horizon, see \cite{Jain:2009uj}.
Our construction of BW potential for hairy black holes might be useful to study
some exotic holographic superconductors with higher order phase instabilities.

The paper is organized as follows. In the next section, we introduce BW construction
by considering the Ising model. We then employ this construction for AdS-Schwarzschild 
and Reissner-Nodstr\"om black hole in order to study the HP transition. This section
is a review of the known results. In section three, we analyze black holes with 
of five dimensional ${\cal{N}} = 2$ gauged supergravity theories with single R-charge.
Besides reproducing the phase diagram given is figure \ref{phased}, we also find 
out various critical exponents near its second order instability line. Section four
is devoted to the study of phase transition involving hairy black holes of 
\cite{Martinez:2005di}. Via BW construction, we critically analyze physics close to the
saddle points representing stable phases. This section also includes an analysis 
of the system under thermal quench. The BW potential can be re-expressed 
in terms of the value of the scalar on the horizon. Treating this as an order
parameter, we construct a {\it time-dependent} solution representing the rolling
of the order parameter from unstable to the stable point after the quench.
In the last section, besides summarizing our
results, we speculate how our results might be useful from the perspective 
of AdS/CFT correspondence.

\section{Bragg-Williams construction: a brief review}

This section is a review of BW theory and pedagogical in nature. It has three subsections.
In the first subsection, we discuss Ising model and use BW theory to capture second order 
paramagnetic to ferromagnetic transition. The later two subsections describe first order 
HP transitions for Schwarzschild and Reissner-Nordstr\"om black holes in AdS space 
respectively.

\subsection{Paramagnetic to ferromagnetic transition}

Bragg-Williams construction is perhaps best described via Ising model \cite{Chaikin}.
Let us consider Ising model on a lattice where, on each site, the classical spin variable $\sigma_{l}$
takes values $\pm 1$. These spins interact via a nearest neighbour coupling $J > 0$.
The Hamiltonian is given by
\begin{equation}
H = - J \sum_{<ll^\prime>} \sigma_l \sigma_{l^\prime}.
\end{equation}
Here the sum is over the nearest neighbour $l$ and $\l^\prime$. The order parameter 
is $m = <\sigma>$, the average of the spin. For spatially uniform $m$, the entropy
can be computed exactly. The total magnetic moment is 
\begin{equation}
m = \frac{N_{+1} - N_{-1}}{N},
\end{equation}
where $N_{+1}$ and $N_{-1}$ are the total number of $+1$ and $-1$ spins respectively. The 
total number of lattice sites is denoted by $N$. The entropy is the logarithm of 
the number of states and is given by
\begin{equation}
S = {\rm ln} (^NC_{N_{+1}}) = {\rm ln}(^N C_{N(1+m)/2})
\end{equation}
which, for entropy per unit spin, gives 
\begin{equation}
s(m) = \frac{S}{N} = {\rm ln}~2 - \frac{1}{2} (1 + m) {\ln}(1 + m) - \frac{1}{2}(1 -m) {\rm ln} (1 - m).
\end{equation}
In BW theory, the energy $<H>$ is approximated via replacing $\sigma$ by its position independent average 
$m$. 
Thus
\begin{equation}
E = - J \sum_{<ll^\prime>} m^2 = - \frac{1}{2} J N z m^2,
\end{equation} 
where $z$ is the number of nearest neighbours in the lattice. 
One then constructs the BW free energy per spin as
\begin{eqnarray}
f(T, m) &=& \frac{E - T S}{N} \nonumber\\
&=& - \frac{1}{2} J z m^2  - T ~{\rm ln}~2 +  \frac{T}{2} (1 + m) {\ln}(1 + m) + \frac{T}{2} (1 - m) {\rm 
ln} (1 
- m). 
\end{eqnarray}
The BW free energy $f(T, m)$ can be plotted as a function of $m$ for various temperatures. It can be 
checked that, for
$T > Jz$, it has a single minimum at $m =0$. However, for $T < Jz$, two minima 
occurs for non-zero values of $m$ leading to paramagnetic to ferromagnetic transition. 
Critical temperature $(T_c)$ for this second order transition can be found by setting first and second 
derivatives of $f$ to zero with the result $T_c = Jz$. For more details, we suggest the readers to
look at {\cite{Chaikin, kubo}.

\subsection{HP transition: AdS-Schwarzschild black hole}

We can implement similar construction  for AdS black holes. Consider a Schwarzschild black hole in 
$(n+2)$ dimensional AdS space. The metric is given by
\begin{equation}
ds^2 = - V(r) dt^2 + V(r)^{-1} dr^2 + r^2 d\Omega_n^2,
\label{sch}
\end{equation}
with 
\begin{equation}
V(r) = \Big(1 - \frac{M}{r^{n-1}} + \frac{r^2}{ l^2}\Big).
\label{comp}
\end{equation}
Here $M$ is a parameter related to the mass or internal energy of the black hole and $l$ is the 
inverse radius of AdS space. We have set $(n+2)$ dimensional gravitational constant $G_{n+2}$ to one.
The black hole has a single horizon where $g_{tt}$ vanishes. We will identify the horizon radius
as $r_+$. The dimensionless temperature, energy and  entropy densities are give by
\begin{eqnarray}
&&\bar T = l T = \frac{(n +1) \bar r^2 + (n -1)}{4 \pi \bar r},\nonumber\\ 
&&\bar E  = l E = \frac{n (\bar r^{n+1} + \bar r^{n-1})}{16 \pi},\nonumber\\
&&\bar S = \frac{\bar r^{n}}{4}.
\label{para}
\end{eqnarray}
Here $l \bar r = r_+$. Before constructing the BW free energy, we will have to
decide on an order parameter. Noticing the form of the entropy and the energy, it
is only natural to consider $\bar r$ as the order parameter. We will see later
that this order parameter has right behaviour expected from the instability 
associated with this black hole. We are now in a position to construct the BW free
energy $\bar {\cal F} (\bar r, \bar T)$ as
\begin{equation}
\bar {\cal F} (\bar r, \bar T) = \bar E - \bar T \bar S = \frac{n (\bar r^{n+1} + \bar r^{n-1})}{16 \pi} 
- \bar T \frac{\bar r^{n}}{4}.
\label{hpfree}
\end{equation}
A plot of the free energy in five dimensions as a function of $\bar r$ for various temperatures is shown 
in figure \ref{hpage}. 
\begin{figure}[t]
\begin{center}
\begin{psfrags}
\psfrag{b}[][]{$\bar {\cal{F}}$}
\psfrag{a}[][]{$\bar r$}
%\psfrag{c}[][]{$\bar T = \frac{1}{\pi}$}
%\psfrag{d}[][]{IV}
%\psfrag{3/4}[][]{$3/4$}
%\psfrag{0.03}[][]{$0.03$}
\epsfig{file=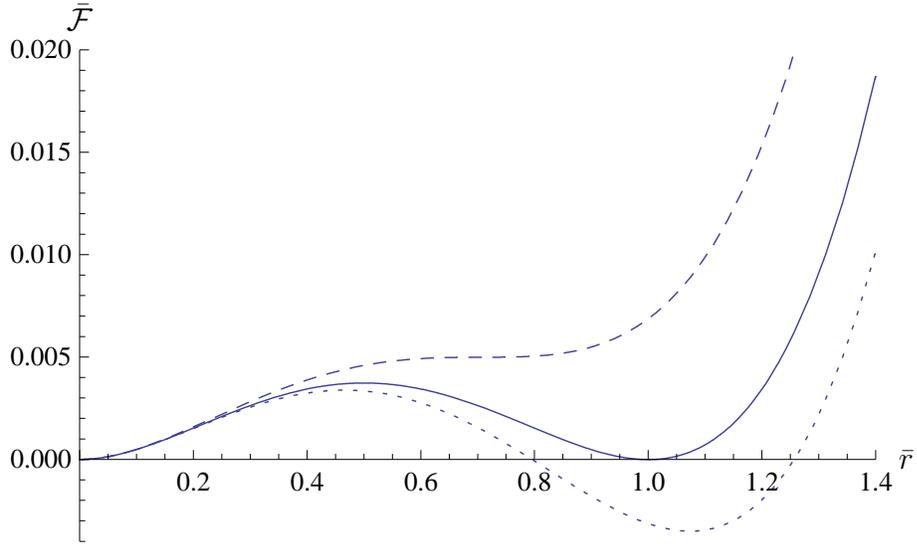, width= 12cm,angle=0}
\end{psfrags}
\vspace{ .1 in }
\caption{BW free energy for five dimensional AdS-Schwarzschild black holes plotted against horizon radius 
$\bar r$ for different temperatures $\bar T$. The solid line has two degenerate minima - representing 
co-existence of black hole phase (minimum at $\bar r =1$) and the thermal AdS phase (with $\bar r = 0$).
This happens at a critical temperature $\bar T_c = 3/(2 \pi)$. While above this temperature black hole
is stable (dotted line), AdS is a preferred phase below $\bar T_c$ (dashed line).  }
\label{hpage}
\end{center}
\end{figure}
Note that in (\ref{hpfree}), the temperature is a parameter. Its dependence on $\bar r$ as given 
in (\ref{para}) appears after minimizing $\bar {\cal{F}}$ with respect to $\bar r$. At this minimum
 $\bar {\cal{F}}$ reduces to the on-shell free energy of the black hole. It is given by
\begin{equation}
\bar F = \bar {\cal{F}}|_{\rm min} = - \frac{\bar r^{n-1} (\bar r^2 -1)}{16 \pi}.
\end{equation}
We identify the AdS free energy with $\bar r$ equals to zero. The first order transition appears 
when 
\begin{equation}
\bar {\cal{F}} = 0, ~{\rm and}~~ \frac{\partial \bar {\cal{F}}}{\partial \bar r} = 0,
\label{condi}
\end{equation}
are satisfied simultaneously. This happens for
\begin{equation}
\bar r = 1, ~{\rm and} ~~\bar T_c = \frac{3}{2 \pi}.
\end{equation}
Below this temperature, black hole phase becomes unstable. As can be seen from the dashed line of 
figure \ref{hpage}, the $\bar r =0$ phase is preferred. This is identified as the AdS phase.
This is a first order phase transition causing a discontinuous change in the order parameter $\bar r$.

\subsection{HP transition: Reissner-Nodstr\"om}

A similar analysis can be performed for charged black holes in the AdS space.
More specifically, we consider here the Reissner-Nodstr\"om black holes. Our
discussion is a brief review of \cite{Banerjee:2010ng}. The metric has the 
same form as (\ref{sch}) with $V(r)$ given by
\begin{equation}
V(r) = \Big(1 - \frac{M}{r^{n-1}} + \frac{q^2}{r^{2n -2}} + {\frac{r^2}{ l^2}}\Big),
\end{equation}
where the additional $q$ dependent term is due to the electric charge that is carried by the 
configuration. The largest root of the equation $V(r) = 0$ represents the outer horizon 
and as before we parametrize it by $r_+$. The chemical potential $\mu$ conjugate to
the charge is given by
\begin{equation}
\mu = \frac{q}{c r_+^{n-1}}, ~{\rm with} ~~c = {\sqrt{\frac{2(n-1)}{n}}}.
\end{equation}
The temperature, energy, entropy and charge  densities are
\begin{eqnarray}
&&\bar T = \frac{(n-1) (1 - c^2 \bar \mu^2) + (n +1) \bar r^2}{4 \pi \bar r},\nonumber\\
&&\bar E = \frac{n}{16 \pi} \Big(\bar r^{n-1} + \frac{q^2}{\bar r^{n-1}} + \bar r^{n+1}\Big),\nonumber\\
&&\bar S = \frac{\bar r^{n-1}}{4}, \nonumber\\
&&\bar Q = {\sqrt{2 n (n-1)}} q.
\label{rnt}
\end{eqnarray}
Here quantities with bars are made dimensionless by multiplying appropriate  factors of $l$ 
wherever
necessary. We have, as before, taken $G_{n+2} = 1$. The BW free energy density, in the grand canonical 
ensemble, is then
\begin{eqnarray}
\bar {\cal{F}}(\bar r, \bar T, \bar \mu) && = \bar E - \bar T \bar S - \bar Q \bar \mu\nonumber\\
&& = n \bar r^{n-1} (1 - c^2 \bar \mu^2) - 4\pi \bar r^n \bar T + n\bar r^{n+1}.
\end{eqnarray}
Qualitative behaviour of the free energy is determined by whether $\bar\mu$ is less than or greater than 
$1/c$. We will only consider the case $\bar\mu < 1/c$. The other case can be found in 
\cite{Chamblin:1999tk}. For fixed $\bar \mu$, one gets a similar graph as in figure \ref{hpage}.
At the saddle point of $\bar {\cal{F}}$, we get 
\begin{equation}
\bar r = \frac{4 \pi \bar T + {\sqrt{ 16 \pi^2 \bar T^2 - 4 (n-1)(1 - c^2\bar \mu^2)}}}{2 (n+1)},
\end{equation}
which, when inverted to get the temperature,  reproduces the one in (\ref{rnt}). 
Critical temperature can be found using (\ref{condi}) with the result
\begin{equation}
\bar T_c = \frac{n \sqrt{1 - c^2 \bar \mu^2}}{2 \pi}.
\end{equation}

Having discussed an application of BW formalism in the study of black hole 
instabilities, in the next section, we discuss another  class of black holes
which show both first and second order instabilities as $(\bar T, \bar \mu)$ are
tuned and we analyze the system within the above framework.

\section{$R$-charged black hole with spherical horizon: Instabilities}

Let us start by briefly recapitulating the black hole 
in five dimensional ${\cal{N}} = 2$ gauged supergravity. Five dimensional 
${\cal{N}} = 2$ gauged supergravity is obtained by compactification of ten dimensional IIB supergravity on 
$S^5$. As shown in \cite{Behrndt:1998jd}, this theory admits asymptotically AdS black hole solutions with 
three $U(1)$ charges with three different horizon topology. For the purpose of this
note we will focus on singly charged black hole with spherical horizon.

The black hole metric with a single $U(1)$ charge is given by
\begin{equation}
ds^2 = - H^{-\frac{2}{3}} f dt^2 +  H^{\frac{1}{3}}
\Big( f^{-1} dr^2 + r^2 d\Omega_3^2\Big),
\end{equation}
where
\begin{equation}
f = 1 - \frac{m}{r^2} + \frac{r^2}{l^2} H, ~~H = 1 + \frac{q}{r^2}.
\end{equation}
In the above equation, $d\Omega_3^2$ is the metric on unit three sphere,
$l$  and $m$ are related to the cosmological constant and the ADM mass of the 
black hole. In particular, $l$ has a dimension of length. The zero of $f$ gives the 
location of the horizon and in the
above parametrization, the horizon appears at $r = r_+$ where
\begin{equation}
r_+ = \Bigg(\frac{-l^2 - q + {\sqrt{(l^2 + q)^2 + 4 m l^2}}}{2}\Bigg)^{\frac{1}{2}}.
\end{equation}
There is a non-trivial gauge field potential associated with this geometry and is given by
\begin{equation}
A_t^i = \frac{{\sqrt{q (r_+^2 + q) (1 + r_+^2)}}}{r^2 + q}.
\end{equation}
From the above we see that $q$ is related to the physical charge. More explicitly,
the physical charge
\begin{equation}
Q = {\sqrt{q (r_+^2 + q) (1 + 
r_+^2)}}.
\label{charge}
\end{equation}
The chemical potential is defined as the value of $A_t^i$ at the 
horizon and is given by
\begin{equation}
\mu = \sqrt{\frac{q (1 + r_+^2)}{r_+^2 + q}}.
\label{chemical}
\end{equation}
It will be convenient for us to scale all the dimensionful quantities
with appropriate powers of $l$ and make them dimensionless. We write all 
these parameters with a bar on the top. For example, the dimensionless 
horizon radius and Hawking temperature of the black hole
are given by,
\begin{equation} 
\bar r = \frac{r_+}{l}, ~~\bar q = \frac{q}{l^2}, ~~\bar T = l T = \frac{2 \bar r^2 + \bar 
q + 1}{2 \pi {\sqrt{\bar r^2 + \bar q}}}.
\label{barred}
\end{equation}
Furthermore, we define the dimensionless Newton's constant $\bar G$ as 
$\bar G = l^3 G$ and set $\bar G = \pi/4$. With this convention, energy and entropy
are given by
\begin{equation}
\bar E = \frac{3}{2} \bar m + \bar q, ~~\bar S = 2 \pi \bar r^2 {\sqrt{\bar r^2 + \bar 
q}}.
\label{rest}
\end{equation}
We would like to study the system in the grand canonical ensemble where we treat 
$\bar T$ and $\bar \mu$ as external parameters. The free energy
is given by
\begin{equation}
\bar F = \bar E - \bar T \bar S - \bar \mu \bar Q = -\frac {\bar r^2 (\bar r^4 + \bar 
\mu^2
-1)}{2 (\bar r^2 - \bar \mu^2 +1)} = -\bar P.
\label{onfree}
\end{equation}
Here $\bar P$ is the pressure. Let us note that $\bar F$ changes sign when $\bar r^4 + 
\bar \mu^2 -1$ changes sign. This is a first order transition and it leads to a crossover 
from AdS phase to the black hole phase. For the gauge theory this represents the 
deconfining transition. Given all these thermodynamic quantities, it is straightforward to 
compute the specific heat and susceptibility. These are given respectively by
\begin{eqnarray}
\bar C &=& \Bigg(\bar T \frac{\partial \bar S}{\partial \bar T}\Bigg)_{\bar \mu} = 
\frac{ 2 \pi \bar r^2 ( 1 + 2 \bar r^2 + \bar q)(3 + 3 \bar r^2 - \bar q) {\sqrt{\bar r^2 
+ \bar q}}}{2 \bar r^4 + \bar r^2 + \bar q \bar r^2 - \bar q^2 + 2 \bar q - 1},
\nonumber\\
\bar \chi &=& \Bigg(\frac{\partial \bar Q}{\partial \bar \mu}\Bigg)_{\bar T} = 
\frac{ (\bar r^2 + \bar q) ( 2 \bar r^4 + \bar r^2 + 5 \bar r^2 \bar q + 6 \bar q - \bar 
q^2 
-1)}{2 \bar r^4 + \bar r^2 + \bar q \bar r^2 - \bar q^2 + 2 \bar q - 1}.
\end{eqnarray}
We note that specific heat and susceptibility diverge at
\begin{equation}
2 \bar r^4 + \bar r^2 + \bar q \bar r^2 - \bar q^2 + 2 \bar q - 1 = 0.
\label{inline}
\end{equation}
This represents the line of continuous phase transition. As one approaches this critical 
line, correlation length diverges. This shows up, as above, in the divergences of
some thermodynamic 
quantities. Near this critical line, the black holes are expected to exhibit some 
universal features. These are encoded in a set of critical exponents normally
called $\alpha, \beta, \gamma$ and $\delta$.
Going close to this line with $\bar \mu$ fixed, we define exponents 
$\alpha, \beta, \gamma$ as 
\begin{equation}
\bar C \sim (\bar T - \bar T_c)^{-\alpha},~~\bar Q - \bar Q_c \sim (\bar T - \bar 
T_c)^{\beta}, ~~\bar \chi \sim  (\bar T - \bar T_c)^{-\gamma}.
\end{equation}
Here $\bar T_c$ is the value of the critical temperature for the chosen $\bar \mu$
(The critical line can be expressed in terms of $\bar T$ and $\bar \mu$ and 
is given later, see (\ref{sorder})). Similarly, one defines $\bar Q_c$. 
The other static exponent $\delta$ is defined as 
\begin{equation}
\bar Q - \bar Q_c \sim (\bar \mu - \bar
\mu_c)^{\frac{1}{\delta}}.
\end{equation}
Here one approaches the critical line with a trajectory on which $\bar T$
is constant. For the black holes in consideration, these quantities are easily 
calculable and are given by
\begin{equation}
\Big(\alpha, \beta, \gamma, \delta\Big) = \Big(\frac{1}{2}, \frac{1}{2}, \frac{1}{2}, 
2\Big).\label{expo1}
\end{equation}
First, note that these exponents are same as those computed for black holes with 
planar horizon \cite{Cai:1998ji,Maeda:2008hn}. Secondly, they satisfy the scaling 
relations
\begin{equation}
\alpha + 2 \beta + \gamma = 2, ~~\gamma = \beta (\delta -1).
\end{equation}

Our main task is now to construct an effective potential that captures all the
phases that we have just discussed. We will use the BW
approach for this purpose. This approach requires 
us to identify an order parameter. Noting the fact that, for a first order transition, 
the change in order parameter is discontinuous and for second order, it changes 
continuously, we continue to use the horizon radius $\bar r$ of the black hole as the order 
parameter.
Once a suitable order parameter is identified, one constructs the BW 
potential which depends on the order parameter, the temperature and the chemical 
potential. This is given by
\begin{equation}
\bar {\cal F} (\bar r, \bar T, \bar \mu) = \bar E - \bar T \bar S - \bar \mu \bar Q.
\end{equation}
In our case, using (\ref{charge}) and  (\ref{rest}), we immediately get\footnote{For 
a black hole with flat horizon a similar construction was provided in \cite{Jain:2009uj}.}
\begin{equation}
\bar {\cal F}(\bar r, \bar T, \bar \mu) = \frac{1}{2} \bar r^2 \Bigg[ 3 - 4 \pi \bar 
T \frac{\bar r 
{\sqrt{1 + \bar r^2}}}
{\sqrt{1 + \bar r^2 - \bar \mu^2}} + r^2 \Big(3 + \frac{\bar \mu^2}{1 + \bar r^2 - \bar 
\mu^2}\Big)\Bigg].
\label{bw}
\end{equation}
The saddle point of $\bar {\cal F}$, namely
\begin{equation}
\frac{\partial \bar {\cal F}}{\partial \bar r} = 0
\label{saddle}
\end{equation}
gives the equilibrium temperature. Using (\ref{bw}), from (\ref{saddle})
we get 
\begin{equation}
\bar T = \frac{\sqrt{1 + \bar r^2}(1 + 2 \bar r^2 - \bar \mu^2)}
{2 \pi \bar r {\sqrt{1 + \bar r^2 - \bar \mu^2}}}.
\label{tempmu}
\end{equation}
Upon using (\ref{chemical}), the above expression 
reduces to the one in (\ref{barred}). Furthermore, substituting (\ref{tempmu})
in (\ref{bw}), we get the on-shell free energy expression as in (\ref{onfree}).
We now proceed to study $\bar {\cal F}$ as we change $\bar T$ and $\bar \mu$.
From the expression of temperature, it is easy to note that it has a minimum
$\bar T_0 = 1/\pi$ when $\bar r = 0$ and $\bar \mu =1$. In what follows, we will focus
ourselves in the region $\bar T \ge \bar T_0$ and $\bar \mu \ge 0$. As noted before, the 
first order transition line is given by the equation
\begin{equation}
\bar r^4 + \bar \mu^2 - 1 = 0.
\end{equation}
Expressed in terms of $\bar T$ and $\bar \mu$, this equation reduces 
to 
\begin{equation}
\bar T = \frac{2+\sqrt{1-\bar \mu^2}}{2 \pi},
\label{forder}
\end{equation}
represented by the dotted line in figure \ref{phased}. 
On the other hand, the second order instability line (\ref{inline}) reads as 
\begin{equation}
%T = \frac{\Delta + \Gamma - \mu^2 \Delta}{\pi \sqrt{\Gamma}}
%\sqrt{ \frac{\Gamma + 2 \Delta}{2 \Delta + \Gamma - 2 \mu^2 \Delta}},
%\label{sorder}
\bar T=\frac{\left(-\Delta  \bar \mu ^2+\Gamma +\Delta \right)}{\sqrt{2} \pi  \Gamma }\sqrt{\frac{\Gamma  (\Gamma +2 \Delta )}{\Delta  \left(\Gamma -2 \Delta  \left(\bar \mu ^2-1\right)\right)}}
\label{sorder}
\end{equation}
where
\begin{equation}
\Delta = (\bar \mu^2 - 1)^{\frac{2}{3}} (\bar \mu + 1)^{\frac{2}{3}}, ~~~
\Gamma = \bar \mu^4 + (\Delta -2) \bar \mu^2 + \Delta^2 - \Delta + 1.
\end{equation}
This is denoted by the solid line in figure \ref{phased}.

To see that $\bar{\cal{F}}(\bar r, \bar T, \bar \mu)$ captures the whole phase
diagram, we first fix $\bar \mu$ and plot $\bar{\cal{F}}$ for various 
temperatures starting with $\bar T = \bar T_0 = 1/\pi$. We start with $\bar \mu =1$. 
The behaviour is 
shown in figure \ref{fixedmu1}. 
\begin{figure}[!]
\begin{center}
\begin{psfrags}
\psfrag{b}[][]{$\bar {\cal F}$}
\psfrag{a}[][]{$\bar r$}
%\psfrag{c}[][]{III}
%\psfrag{d}[][]{IV}
%\psfrag{3/4}[][]{$3/4$}
%\psfrag{0.03}[][]{$0.03$}
\epsfig{file=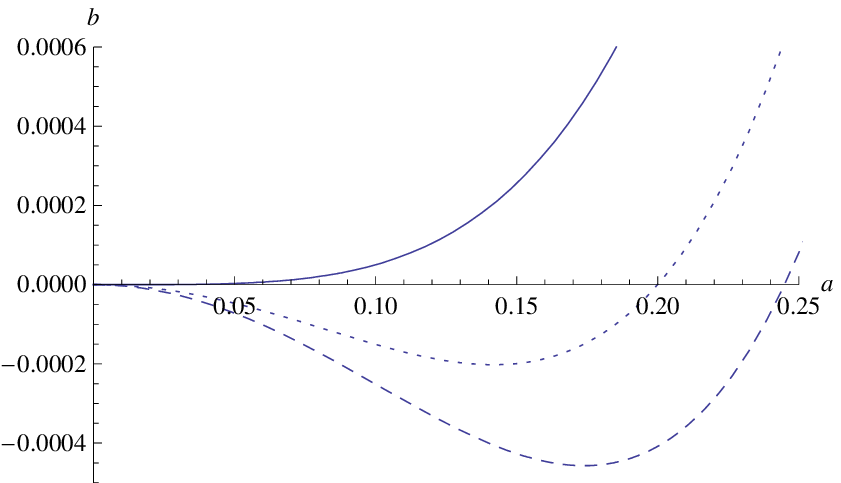, width= 7cm,angle=0}
\end{psfrags}
\vspace{ .1 in }
\caption{$\bar {\cal F}$  is plotted as a function of the order parameter $\bar r$ for 
$\bar \mu = 1$. 
The solid, dotted and dashed curves are for $\bar T = 1/\pi, 1.01/\pi, 1.015/\pi$.}
\label{fixedmu1}
\vspace{.1 in}
\begin{psfrags}
\psfrag{b}[][]{$\bar {\cal F}$}
\psfrag{a}[][]{$\bar r$}
\epsfig{file=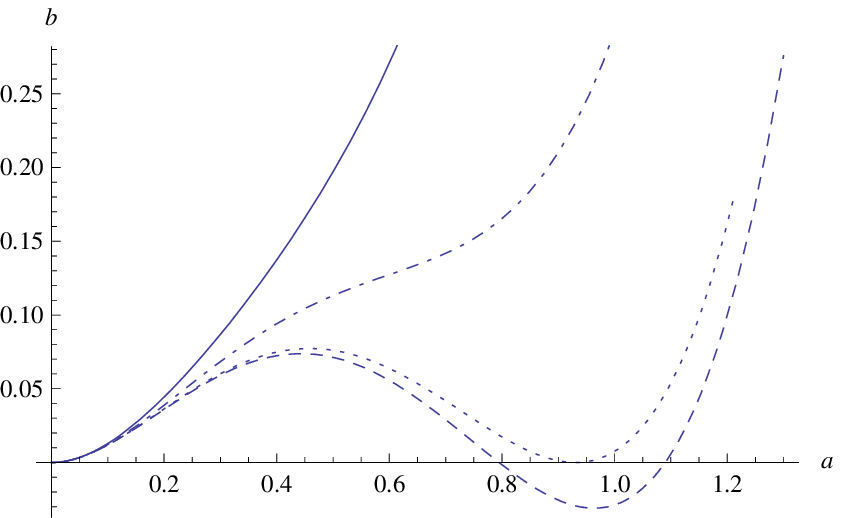, width= 7cm,angle=0}
\end{psfrags}
\caption{$\bar {\cal F}$  is plotted as a function of the order parameter $\bar r$ for
$\bar \mu = .5$. The solid, dot-dashed, dotted and dashed lines are for
$\bar T = 1/\pi, 1.3/\pi, 1.433/\pi, 1.45/\pi$.}
\label{fixedmuhalfa}
\vspace {.1 in}
\begin{psfrags}
\psfrag{b}[][]{$\bar {\cal F}$}
\psfrag{a}[][]{$\bar r$}
\epsfig{file=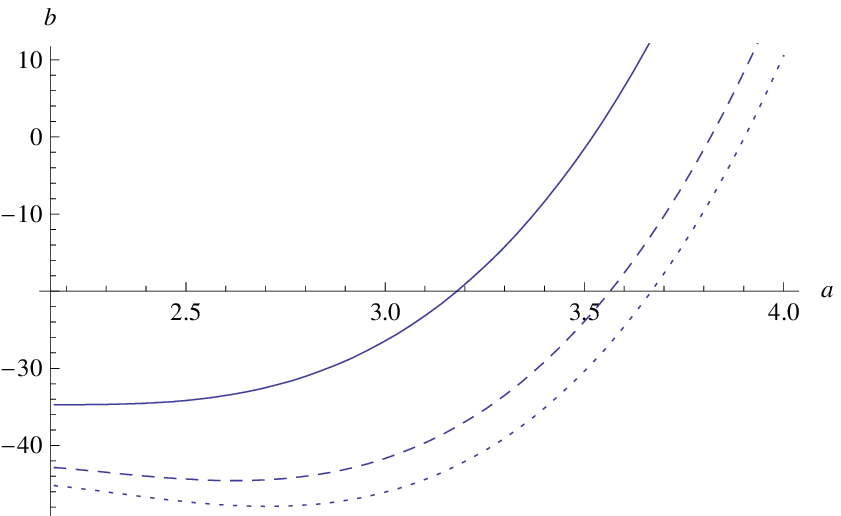, width= 7cm,angle=0}
\end{psfrags}
\caption{$\bar {\cal F}$  for $\bar \mu =2$. Solid, dashed and dotted
lines are for $\bar T = 0.86, 0.93, 0.95$ respectively. Solid line
represents $\cal F$ at critical temperature. Below this temperature,
black hole becomes unstable. The minima in the rest two curves show
the stable black hole phase.}

\label{mutwocurve}
\end{center}
\end{figure}
We note that at $\bar T = \bar T_0 = 1/\pi$, {$\bar {\cal F}$ has a minimum
at $\bar r = 0$. Its first and second derivatives with respect to $\bar r$ also vanish 
at that point. In this sense, it is a point of inflection for $\bar {\cal F}$. If we 
increase $\bar T$ further, we get minima for increasing values of $\bar r$ --  
representing stable black hole phases with increasing size. This is in complete 
agreement with the phase diagram in figure \ref{phased}. Next, we analyze the system 
for $0 \le \bar \mu <1$. From figure \ref{phased}, we expect that $\bar {\cal F}$ 
should show a HP transition as we increase the temperature beyond a critical value.
We precisely see this in figure \ref{fixedmuhalfa}, where we  have plotted $\bar {\cal 
F}$ for $\bar \mu = .5$. While the point $\bar r = 0$ is identified with
the AdS phase, any finite value of $\bar r$ represents a black hole with $\bar r$ 
being the horizon. As we increase the temperature, we note a crossover from 
AdS to the black hole phase at $\bar T = \bar T_{HP} = 1.433/\pi$. This is shown by the 
dotted line in the figure. At this temperature the order parameter $\bar r$ changes 
discontinuously from zero to a finite value - clearly a signature of a first-order 
transition. 
%\begin{figure}[t] 
%\begin{center}
%\begin{psfrags}
%\psfrag{b}[][]{$\bar {\cal F}$}
%\psfrag{a}[][]{$\bar r$}
%\psfrag{c}[][]{III}
%\psfrag{d}[][]{IV}
%\psfrag{3/4}[][]{$3/4$}
%\psfrag{0.03}[][]{$0.03$}
%\epsfig{file=muhalffixed.eps, width= 12cm,angle=0}
%\end{psfrags}
%\vspace{ .1 in }
%\caption{$\bar {\cal F}$  is plotted as a function of the order parameter $\bar r$ for
%$\bar \mu = .5$. The solid, dot-dashed, dotted and dashed lines are for 
%$\bar T = 1/\pi, 1.3/\pi, 1.433/\pi, 1.45/\pi$.}
%\label{fixedmuhalfa}
%\end{center}
%\end{figure}
Now as we decrease $\bar \mu$, HP transition temperature increases. In particular,
for $\bar \mu = 0$, $\bar T_{HP} = 3/(2 \pi)$ as expected for AdS-Schwarzschild black 
hole.  Finally, we increase $\bar \mu$ beyond $1$. For $\bar \mu =2$, 
$\cal F$ is shown in figure \ref{mutwocurve}. Plots are shown for 
different temperatures, starting with the critical one (solid curve). Below 
this temperature, we reach the yet unknown phase and the black hole is
unstable. At higher temperatures (dashed and dotted curve), minima 
of the curves represent the stable black hole phases. 

We can continue the same exercise for $\bar T$ fixed at any value above
$1/\pi$ and change $\bar \mu$. For $\bar T_{HP} \le 3/(2 \pi)$ and $\bar \mu \le 1$,
we first cross the HP line. Close to this point, $\cal F$ behaves similar to 
 that of figure \ref{fixedmuhalfa}. Further increasing $\bar \mu$ but keeping $\bar T$ 
fixed, we 
hit the continuous phase transition line leading to figure \ref{mutwocurve}. 
For $\bar T \ge 3/(2 \pi)$, the first order transition is lost. Black hole
is always a stable phase for low $\bar \mu$. However, as we take $\bar \mu$ to
a critical value, black hole ceases to be stable and we reach the second order line
getting a figure similar to  figure \ref{mutwocurve}.

Finally, let us now discuss about the procedure for obtaining the critical exponents from the mean 
field 
potential $\bar{\mathcal F}$ which has already been written in (\ref{expo1}). We note that the specific heat at fixed chemical potential can be obtained from (\ref{bw}).
\begin{equation}
 \bar C_{\bar \mu}=-\bar T\frac{\partial^2 \bar{\mathcal F} }{\partial \bar T^2}\Bigg{|}_{\bar\mu}\sim (\bar T-\bar T_c)^{-\frac{1}{2}},
\end{equation} 
which gives $\alpha=\frac{1}{2}$. If we approach the critical line along constant $\bar \mu=\bar \mu_c$, then we see that
\begin{equation}
 \bar Q - \bar Q_c \sim (\bar T -\bar T_c)^{\frac{1}{2}},
\end{equation}
where $\bar Q_c$ is the critical value of the charge at fixed $\bar \mu_c$. This shows that the critical exponent $\beta$ has the value $\frac{1}{2}$. Similarly, the susceptibility behaves near the critical temperature as 
\begin{equation}
 \chi= \frac{\partial \bar Q}{\partial \bar \mu}\Bigg{|}_{\bar T}\sim (\bar T -\bar T_c)^{-\frac{1}{2}}.
\end{equation}
This leads us to the critical exponent $\gamma=\frac{1}{2}$. Finally, on approaching the critical line with $\bar T=\bar T_c$ we get
\begin{equation}
 \bar Q -\bar Q_c \sim (\bar \mu -\bar \mu_c)^{\frac{1}{2}}.
\end{equation}
So, this gives us $\delta=2$.

\section{Hairy to Reissner-Nodstr\"om black holes: a continuous phase transition}

In this section, we first review the main features of the hairy black holes \cite{Martinez:2005di}
and their instability. We then characterize this instability via BW construction
and argue that this black holes undergo a continuous transition at high temperature.

We consider four dimensional gravity action in the presence of a negative cosmological constant
where the matter content is given by a conformally coupled real self interacting scalar field
and a Maxwell gauge field. 
\begin{equation}
S = \int d^4 x {\sqrt{-g}}\Bigg( \frac{1}{16\pi} \Big(R + \frac{3}{l^2}\Big) - 
\frac{F_{\mu\nu}F^{\mu\nu}}{16\pi} - \frac{1}{2}g^{\mu\nu} \partial_\mu\phi \partial^\mu\phi -
\frac{1}{12} R \phi^2 - \alpha \phi^4\Bigg).
\label{conform}
\end{equation}
The black holes of this theory are described by the metric
\begin{equation}
ds^2 = - V(r) dt^2 + V(r)^{-1} dr^2 + r^2 d\sigma^2,
\end{equation}
with
\begin{equation}
V(r) = \frac{r^2}{l^2} - \Bigg(1 + \frac{M}{r}\Bigg)^2.
\end{equation}
In the expression of the metric, $d\sigma^2$ represents the line element of a constant
negative curvature two dimensional manifold. The scalar and the non-zero component of the electromagnetic 
field are given by
\begin{equation}
\phi = {\sqrt{ \frac{1}{2\alpha l^2}}} \Bigg(\frac{M}{r +  M}\Bigg),
~~A_t(r) = -\frac{q}{r}.
\label{phi}
\end{equation}
It is important to note that the mass and charge are not independent but related via
\begin{equation}
q^2 = M^2 \Bigg(\frac{2 \pi }{3 l^2 \alpha} -1 \Bigg) =  M^2  (a -1).
\label{q-m}
\end{equation}
Here $a$ is defined as
\begin{equation}
a = \frac{2 \pi }{3 l^2 \alpha}.
\label{def}
\end{equation}
In terms of appropriately scaled variables, the temperature, chemical potential, internal energy, charge,  
and entropy densities are given by \cite{Martinez:2005di}
\begin{eqnarray}
&&\bar T = \frac{1}{2\pi} (2 \bar r -1), ~~\bar \mu = \frac{\bar q}{\bar r},\nonumber\\
&&\bar E = \frac{1}{4\pi} \bar r (\bar r -1), ~~\bar Q = \frac{\bar q}{4\pi}, \nonumber\\
&&\bar S = \frac{\bar r^2}{4} \Big(1 - \frac{a (\bar r - 1)^2}{\bar r^2}\Big).
\label{thermo}
\end{eqnarray}
Note that due to the conformal coupling of the scalar to the curvature, the entropy 
density gets modified from standard form by an``effective" gravitational constant
\cite{Martinez:2010ti}. We also note that entropy remains positive only in the 
temperature range
\begin{equation} 
\frac{1}{2\pi}  \Bigg(\frac{\sqrt{a} -1}{\sqrt{a} + 1}\Bigg) \le \bar T \le \frac{1}{2\pi} 
\Bigg(\frac{\sqrt{a} 
+ 1}{\sqrt{a} - 1}\Bigg).
\label{range}
\end{equation} 
We call the limiting values to be $\bar T_{\rm min}, \bar T_{\rm max}$ respectively.

There is an additional black hole solution to the action (\ref{conform}). We will
call this the Reissner-Nordstr\"om solution. The metric has the 
form \cite{Martinez:2005di}
\begin{equation}
ds^2 = - V(\rho) dt^2 + V(\rho)^{-1} d\rho^2 + \rho^2 d\sigma^2,
\end{equation}
with
\begin{equation}
V(\rho) = \frac{\rho^2}{l^2} - \Bigg(1 + \frac{2M_0}{\rho} - \frac{q_0^2}{\rho^2} \Bigg).
\end{equation}
with 
\begin{equation} 
\phi = 0, ~{\rm and} ~~A_t = -\frac{q_0}{\rho}.
\end{equation}
The event horizon is located at $V(\rho) = 0$, the solution of which we will call $\rho_+$.
Thermodynamic quantities associated with this black holes are
\begin{eqnarray}
&&\bar T = \frac{1}{2\pi} \Bigg( \frac{3}{2} \bar \rho - \frac{1}{2\bar \rho} - 
\frac{\bar q_0^2}{2 \bar \rho^3}\Bigg),\nonumber\\
&&\bar E = \frac{1}{8\pi} \Bigg(\bar \rho^3 - \bar \rho - \frac{q_0^2}{\bar \rho}\Bigg),\nonumber\\
&&\bar Q = \frac{\bar q_0}{4 \pi}, ~~\bar S = \frac{\bar \rho^2}{4}, ~~\bar \mu = \frac{\bar q_0}{\bar \rho}.
\end{eqnarray}

In the following, we will argue that the hairy black holes, in the grand canonical ensemble, are unstable
and crosses over to the RN black holes at high temperature. We will also characterize this instability via BW 
analysis. First of all, in order to compare two different black holes, namely the RN and the hairy one, 
we will have to make sure that they have the same temperature and chemical potential.
That means 
\begin{eqnarray}
\frac{1}{2\pi} \Big( \frac{3}{2} \bar \rho - \frac{1}{2\bar \rho} -
\frac{\bar q_0^2}{2 \bar \rho^3}\Big) &=& \frac{1}{2\pi} (2 \bar r -1),\nonumber\\
\frac{\bar q_0}{\bar \rho} &=& \frac{\bar q}{\bar r}.
\label{equate}
\end{eqnarray}
These two equations allow us to express $\bar q_0$ and $\bar\rho$ in terms of
$\bar q$ and $\bar r$. In particular, for $\bar\rho$, we get
\begin{equation}
\bar \rho = \frac{1}{3\bar r} \Big(- \bar r + 2 \bar r^2 + \sqrt{3 \bar q^2 + 4 \bar r^2 - 4 \bar r^3 + 4\bar 
r^4}\Big).
\label{rhor}
\end{equation}

The BW free energy density for both the black holes can now be easily computed as was done in the 
previous sections. For the hairy one it reads
\begin{eqnarray}
4 \pi \bar{\cal{F}}_{\rm hair} &=& 4 \pi (\bar E - \bar T \bar S - \bar Q \bar 
\mu)\nonumber\\
&=& \bar r (\bar r - 1) - \pi \bar r^2 \Big(1 - a \frac{(\bar r -1)^2}{\bar r^2}\Big) \bar T - 
\bar r (\bar r -1) \sqrt{a -1} ~\bar \mu\nonumber\\
&=& \bar r (\bar r - 1) - 
\pi \bar r^2 \Big(1 - a \frac{(\bar r -1)^2}{\bar r^2}\Big) \bar T - \frac{(a - 
1)}{2} 
\bar r (\bar r -1) (2 \pi \bar T -1).
\label{fhair}
\end{eqnarray}
In going from the first line to the second, we use the fact that for hairy black holes, $\bar q$ is 
not independent, but related to $\bar \mu$ and hence $\bar r$ through (\ref{q-m}). Similarly,
the conjugates $\bar \mu$ is related to $\bar T$ via
\begin{equation}
\bar \mu = \frac{1}{2} \sqrt{a -1} ( 2 \pi \bar T -1).
\end{equation}
We used this equation to get to the last line of (\ref{fhair}).
As for RN black holes, we can proceed similarly to get
\begin{eqnarray}
4 \pi \bar{\cal{F}}_{\rm RN} &=& 4 \pi (\bar E - \bar T \bar S - \bar Q \bar
\mu)\nonumber\\
&=& \frac{\bar \rho^3}{2} - \frac{\bar \rho}{2} + \frac{q_0^2}{2 \bar \rho} - \pi \bar \rho^2 \bar T
- \bar q_0 \bar \mu\nonumber\\
&=&  \frac{\bar \rho^3}{2} - \frac{\bar \rho}{2} + \frac{\bar\rho}{2}(a-1) (\bar r - 1)^2
 - \pi \bar\rho^2 \bar T - \frac{\bar \rho}{2}(a -1) (\bar r - 1) (2 \pi \bar T -1),
\label{frn}
\end{eqnarray}
where we need to substitute $\bar \rho$ using (\ref{rhor}) and further $\bar q$ by $\bar m$ and 
hence by $\bar r$. In order to write (\ref{frn}), we have also made use of the second identification 
given in (\ref{equate}).
Further, using (\ref{rhor}) and (\ref{fhair}), after some simplification, we can re-write
$\bar{\cal{F}}_{\rm RN}$ as 
\begin{eqnarray}
\bar{\cal{F}}_{\rm RN} (\bar r, \bar T, \bar a) &=&
\frac{1}{54} \Big[ (1 + \bar r) (-1 - \bar r + \delta) (-5 + 4 \bar r - 6\pi \bar T)
+ 3 a (\bar r -1) \{3 + 12 \bar r^2 - \delta \nonumber\\
&&+~30 \pi \bar T - 6 \pi \delta \bar T 
+ \bar r (-21 +4 \delta - 18 \pi \bar T)\}\Big] + \bar r (\bar r - 1)  \nonumber\\
&&- \pi \bar r^2 \Big(1 - a \frac{(\bar r -1)^2}{\bar r^2}\Big) \bar T - \frac{(a -
1)}{2}
\bar r (\bar r -1) (2 \pi \bar T -1).
\label{freedif}
\end{eqnarray}
with 
\begin{equation}
\delta = \sqrt{ 3 a (\bar r -1)^2 + (\bar r + 1)^2}.
\end{equation}
The saddle point of (\ref{fhair}) and (\ref{freedif}) occurs at
\begin{equation}
\bar r = \frac{1}{2} (1 + 2 \pi \bar T),
\label{ons}
\end{equation}
and at the minima, 
\begin{eqnarray}
&&\bar{\cal{F}}_{\rm hair} = -\frac{1}{8\pi} \Big( \bar r^2 + a (\bar r -1)^2 
                                 \Big),\nonumber\\
&&\bar{\cal{F}}_{\rm RN} =
\frac{1}{216\pi} \Big(2 + 6 \bar r - 21 \bar r^2 + 2 \bar r^3 - 2 \delta (1 + \bar r)^2 - 
   3 a (-1 + \bar r)^2 (-3 + 2 \delta + 6 \bar r)\Big).\nonumber\\
\end{eqnarray}
\begin{figure}[!]
\begin{center}
\begin{psfrags}
\psfrag{b}[][]{$\bar {\cal{F}}$}
\psfrag{a}[][]{$\bar r$}
%\psfrag{c}[][]{$\bar T = \frac{1}{\pi}$}
%\psfrag{d}[][]{IV}
%\psfrag{3/4}[][]{$3/4$}
%\psfrag{0.03}[][]{$0.03$}
\epsfig{file=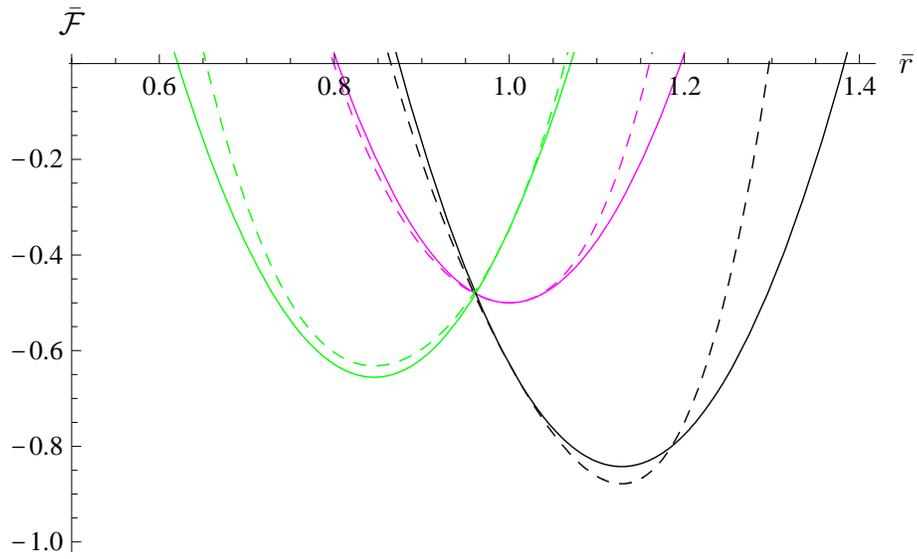, width= 12cm,angle=0}
\end{psfrags}
\vspace{ .1 in }
\caption{This figure is a plot of (\ref{fhair}) and (\ref{freedif}) for $a = 25$
and for different temperatures. The solid lines and the dashed lines represent the 
hairy and the RN black holes respectively. Green, magenta and black curves are
for $\bar T = 0.11, 1/(2 \pi), 0.2$ respectively. We see that while at low temperature
free energy is minimized by the hairy black hole, RN black holes dominate 
at high temperature. At $\bar T = 1/(2 \pi)$, free energies are equal at the minimum.}
\label{n-i}
\end{center}
\end{figure}
While for $\bar T \le \bar T_c = 1/(2 \pi)$, $\bar{\cal{F}}_{\rm hair}$ minimizes the
free energy, for $\bar T \ge \bar T_c$, RN represents the stable black holes.
From (\ref{ons}), it follows that at the critical temperature $\bar T_c$, $\bar r = \bar r_c =1$.
Near $\bar T_c$ it follows that
\begin{equation}
\bar r - \bar r_c \sim \bar T - \bar T_c, ~~~ \bar {\cal F} - \bar {\cal F}_c \sim (\bar T - \bar
T_c)^3,
\label{cx}
\end{equation}
where $ \bar {\cal F}_c$ is the value of $\bar {\cal F}$ at $\bar r = \bar r_c$. The
derivative of specific heat with respect to temperature has a discontinuity around
$\bar r_c$ of $(2 + a)\pi^2$. This is thus a continuous phase transition from hairy
to RN black holes. The critical exponent following from (\ref{cx}) is $\alpha = -1, \beta = 1$. 
In figure \ref{n-i}, we have plotted $\bar {\cal F}$ for
different black holes at different temperatures and scalar couplings. The
behaviour of $\bar {\cal F}$ close $\bar T_c$ is shown in figure \ref{lowa}.
\begin{figure}[!]
\begin{center}
\begin{psfrags}
\psfrag{b}[][]{$\bar {\cal{F}}$}
\psfrag{a}[][]{$\bar r$}
%\psfrag{c}[][]{$\bar T = \frac{1}{\pi}$}
%\psfrag{d}[][]{IV}
%\psfrag{3/4}[][]{$3/4$}
%\psfrag{0.03}[][]{$0.03$}
\epsfig{file=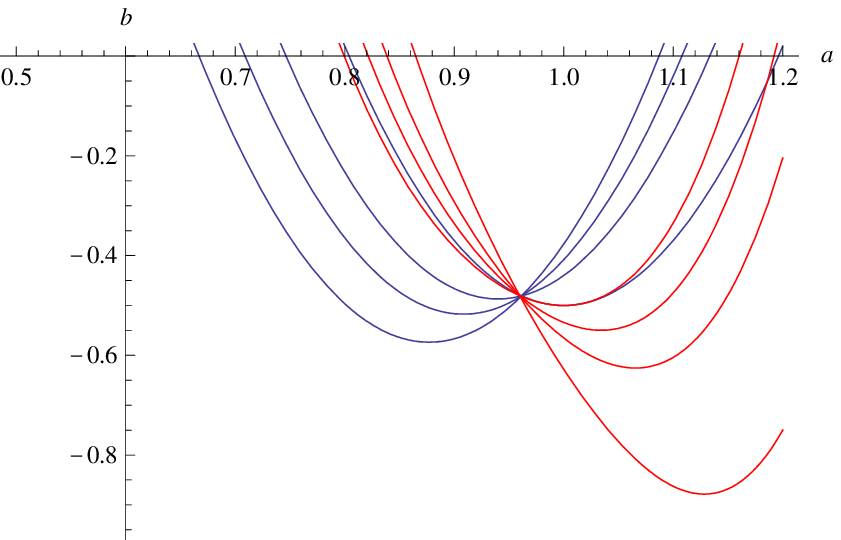, width= 12cm,angle=0}
\end{psfrags}
\vspace{ .1 in }
\caption{ This figure is the behaviour of the free energy function close
to $\bar T = \bar T_c = 1/(2 \pi)$ for $a = 30$. Blue and red are for $\bar T \le T_c$ and 
$\bar T \ge \bar T_c$ representing hairy and RN 
black holes respectively. At $\bar T = \bar T_c$, the minima for both are degenerate. Clearly, the 
order parameter $\bar r$, at which the minima occur, changes continuously around critical
temperature leading to a continuous phase transition.}
\label{lowa}
\end{center}
\end{figure}

We note that the BW free
energy constructed in (\ref{fhair})
can also be expressed using the value of the scalar $\phi$ at the
horizon as order parameter. Inverting (\ref{phi}),
we can express $\bar {\cal F}_{\rm{hair}}$ as,
\begin{equation}
4 \pi \bar{\cal{F}}_{\rm hair} = \frac{\sqrt a}{({\sqrt{\frac{3 a}{\pi}}} - 2 \phi_h)^2}
\Big( 4 {\sqrt a} \pi \bar T \phi_h^2 + {\sqrt{\frac{3}{\pi}}} (1 + a - 2 (a-1) \pi \bar T)\phi_h
- 3 {\sqrt a} \bar T\Big).
\label{bwphi}
\end{equation}
Here $\phi_h$ is the value of the scalar at the horizon.
The expression on the right has a minimum at
\begin{equation}
\phi_h = {\sqrt{\frac{3 a}{4 \pi}}} \Bigg(\frac{2 \pi \bar T -1}{2 \pi \bar T +1}\Bigg),
\label{phihstable}
\end{equation}
such that, for $\bar T = \bar T_c$, $\phi = 0$.

Having reached this far, we now like to address some dynamical issues associated
with this system. In particular, we ask as to how the order parameter $\phi_h$ 
behaves in time as we temperature quench the system from $\bar T > \bar T_c$
to $\bar T < \bar T_c$. We assume that, during the quench, the temperature 
changes so fast that $\phi_h$, immediately after the change, is identical to its 
value before . However, at a later time $\phi_h$ must roll down to its stable 
position given by (\ref{phihstable}). In the following, we will be interested in
finding out the interpolating solution  $\phi_h(t)$ which connects the unstable
to the stable point.

The equation that we need to solve is
\begin{equation}
\partial_t^2 \phi_h(t) + \frac{\partial \bar{\cal{F}}_{\rm hair}}{\partial \phi_h(t)} = 0,
\end{equation}
where $\bar{\cal{F}}_{\rm hair}$ is given by (\ref{bwphi}). This equation can be 
immediately  integrated once to get
\begin{equation}
\frac{1}{2} (\partial_t \phi_h)^2 + \bar{\cal{F}}_{\rm hair} (\phi_h) = C.
\end{equation}
The integration constant $C$ can be fixed by the boundary condition 
$\partial_t \phi_h = 0$ for $\phi_h = 0$. This gives
\begin{equation}
\frac{1}{2} (\partial_t \phi_h)^2 + \bar{\cal{F}}_{\rm hair} (\phi_h) = 
\bar{\cal{F}}_{\rm hair} (0)
\end{equation}
It turns out that this equation can be integrated exactly with the 
boundary condition $\phi_h (t) = 0$ at $t = 0$. The result can 
be expressed in a form 
\begin{equation}
f ( \phi_h, a, \bar T) = t,
\end{equation}
where $f$ is a known function of $\phi_h$. Furthermore,
it has parametric dependences on $\bar T$ and $a$. This function
is too non-illuminating and hence we do not display it here. It however
turns out that the equation above can not be analytically inverted to get
$\phi_h(t)$ as an explicit function of $t$. Nevertheless, numerically it can be solved
and the result is shown in the figure \ref{lowb}.
\begin{figure}[!]
\begin{center}
\begin{psfrags}
\psfrag{f}[][]{$\phi_h(t)$}
%\psfrag{a}[][]{$\bar r$}
%\psfrag{c}[][]{$\bar T = \frac{1}{\pi}$}
%\psfrag{d}[][]{IV}
%\psfrag{3/4}[][]{$3/4$}
%\psfrag{0.03}[][]{$0.03$}
\epsfig{file=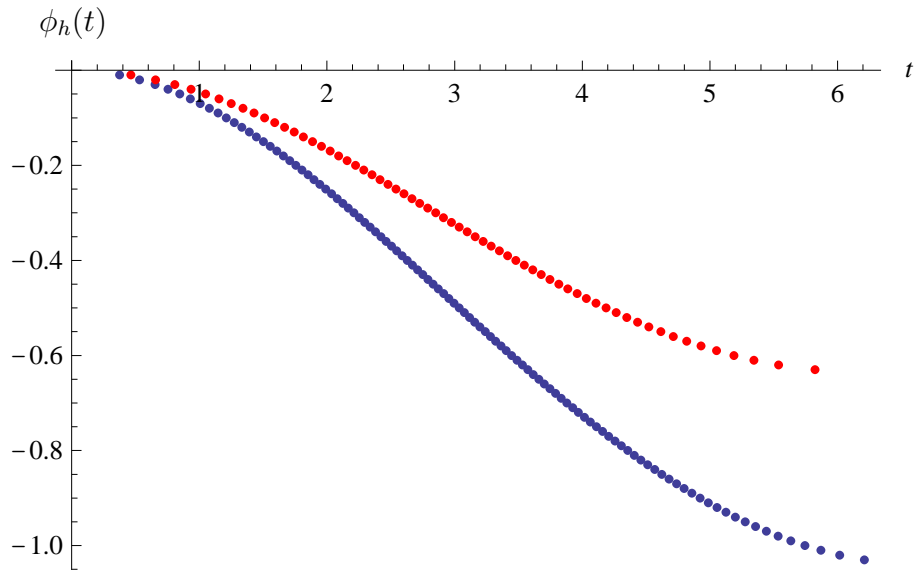, width= 12cm,angle=0}
\end{psfrags}
\vspace{ .1 in }
\caption{ This figure shows behaviour of $\phi_h(t)$ after quenched to different temperatures
below $\bar T_c = 1/(2\pi)$. The vertical axis is $\phi_h$ and the horizontal one is $t$.
The plots are for $a = 90$. While the lower one (blue) curve is for temperature 
quenched to $\bar T = .13$, the upper one (red) is for $\bar T = .14$. We see
$\phi_h(t)$ starts with zero value at $t =0$ and at a later time reaches a non-zero
negative stable point determined by the equation (\ref{phihstable})}.
\label{lowb}
\end{center}
\end{figure}
In the plot, we have shown two cases where temperature $\bar T$ is quenched down
to $.14$ (red) and $.13$ (blue). The value of $a$ that we have chosen is $90$. Starting
from $\phi_h(t)  = 0$ at $t =0$, $\phi_h(t)$ rolls down to respective stable points dictated
by the equation (\ref{phihstable}).

\section{Summary}

In this work our aim was to study black hole instabilities within the
framework of BW theory of phase transition. After providing a pedagogical
review to this subject, we employed BW method in two cases. One involved
the $R$-charged black holes with spherical horizon in five dimensional AdS space. 
In the presence of non-zero chemical potential, it undergoes both first and
second order transitions. We found that BW theory, with horizon radius as 
order parameter, captures all these instabilities. We hope that, 
via AdS/CFT correspondence, the constructed BW free
energy will be useful to study the phases of strongly coupled 
${\cal N} = 4$ SYM theory on $R^3$ at finite temperature and chemical
potential in the same way as in \cite{Jain:2009uj}. 

The other example that we studied is the fate of four dimensional
hairy black holes with hyperbolic horizon. Again, via a BW analysis we 
argued that with the increase in temperature, this black hole 
becomes unstable, loses its ``hair" and turns into a stable RN black hole.
This transition is analogous to a third order phase transition with 
a singularity in the derivative of the specific heat. The BW free
energy is constructed in (\ref{fhair}). Using value of the scalar on
the horizon as order parameter, we studied its behaviour under temperature
quench. The corresponding rolling down solutions were semi-analytically 
constructed. 

Within the AdS/CFT correspondence, in \cite{Gubser:2008px, Hartnoll:2008vx}, 
second order instabilities associated with hairy black holes with flat horizon were used to 
understand holographic superconductors at the boundary. We note that superconductors with possible
higher order transition (similar to the one we discussed) has been reported earlier, see for example
\cite{htk}. We hope a construction like (\ref{bwphi}) will be useful to analyse such holographic
superconductors, however in hyperbolic space. 

\section*{Acknowledgment}

We have benefited from discussions with Swarnendu Sarkar and Goutam 
Tripathy.

\end{document}